\documentclass[aps,twocolumn,prl,showpacs,showkeys,preprintnumbers,superscriptaddress,nobibnotes,floatfix,nofootinbib]{revtex4-2}

\usepackage{orcidlink}
\usepackage{amsmath}
\usepackage{amsfonts}
\usepackage{amssymb}
\usepackage{mathrsfs}
\usepackage{bm} 
\usepackage{bbm} 
\usepackage{slashed} 
\usepackage{physics} 
\usepackage{esvect} 
\usepackage[version=4]{mhchem}
\usepackage{xcolor} 
\usepackage{graphicx} %
\usepackage{array} 
\usepackage{url} %
\usepackage{amsthm} 
\usepackage{enumerate} 
\usepackage{enumitem} %
\usepackage{booktabs} 
\usepackage{siunitx} 
\usepackage{mdframed} %
\usepackage{tcolorbox} %
\usepackage{fontawesome5} 
\usepackage{braket} 
\usepackage{setspace} %
\usepackage{multirow} %
\usepackage{upgreek} 
\usepackage{xspace} %
\usepackage[normalem]{ulem} 
\usepackage{array}
\usepackage{natbib} 
\usepackage{hyperref} %
\hypersetup{
    colorlinks=true,
    linkcolor=blue,
    citecolor=blue,
    filecolor=magenta,
    urlcolor=blue,
    breaklinks=true
}
\usepackage[all]{hypcap} 
\usepackage[capitalise]{cleveref} %

\setlist[description]{leftmargin=\parindent,labelindent=\parindent} %
\newcommand{\prlsection}[2]{{\it\textbf{#1}{#2}}---}

\usepackage{slashed}

\begin{document}

\title{Inelastic Dark Matter Signature at High Recoil Energy in LUX--ZEPLIN and CRESST}%

\author{Liangliang Su}
\email{liangliang.su@kit.edu}
\affiliation{Institute for Astroparticle Physics (IAP), Karlsruhe Institute of Technology (KIT),
Hermann-von-Helmholtz-Platz 1, 76344 Eggenstein-Leopoldshafen, Germany}

\author{Jin Min Yang}
\email{jmyang@itp.ac.cn}
\affiliation{Centre for Theoretical Physics, Henan Normal University, Xinxiang 453007, P. R. China}
\affiliation{Institute of Theoretical Physics, Chinese Academy of Sciences, Beijing 100190, P. R. China}

\author{Wen-Na Yang}
\email{wennayang@njnu.edu.cn}
\affiliation{Department of Physics and Institute of Theoretical Physics, Nanjing Normal University, Nanjing, 210023, P. R. China}

\date{\today}

\begin{abstract}
The LUX--ZEPLIN Collaboration recently reported one event at $E_{\mathrm{nr}}=248\pm23~\mathrm{keV}$ with an exposure of $2.84~\mathrm{tonne\cdot yr}$. In this Letter, we interpret this feature using endothermic dark matter (DM). We consider both direct scattering of the surviving ground-state halo component and the contribution of excited states produced by terrestrial upscattering. Our calculations show that explaining the high-energy event requires $m_\chi\gtrsim500~\mathrm{GeV}$ and a mass splitting of $\mathcal{O}(300)~\mathrm{keV}$, for which the production of excited states inside the Earth is kinematically forbidden. For $\bar{\sigma}_n=10^{-37}~\mathrm{cm^2}$, an illustrative two-bin likelihood analysis yields a representative best-fit point at $(m_\chi,\delta)\simeq(1.105~\mathrm{TeV},350~\mathrm{keV})$. The preferred parameter region may be tested by the planned CRESST upgrade.

\end{abstract}

\maketitle

\prlsection{Introduction}{.} The evidence for dark matter (DM) comes from its gravitational effects on galaxies and galaxy clusters, as well as its imprint on the expansion history of the Universe~\cite{Rubin:1970zza,Clowe:2006eq,Planck:2018vyg}. These observations indicate that DM accounts for roughly one quarter of the Universe’s energy~\cite{Planck:2018vyg}. Although its gravitational effects are well established, its microscopic nongravitational nature remains unknown. In direct detection searches, the velocity distribution of halo DM limits the available kinetic energy, leading most searches to focus on nuclear and electron recoil signals below approximately \(100~\mathrm{keV}\). These include liquid xenon detectors targeting weakly interacting massive particles (WIMPs)~\cite{Lee:1977ua, Jungman:1995df} with masses from a few GeV to several TeV, as well as novel detectors for light DM research~\cite{Essig:2011nj,Essig:2015cda,Kouvaris:2016afs,Knapen:2017xzo,Bringmann:2018cvk,Alvey:2019zaa,Wang:2019jtk,Ge:2020yuf,Guo:2020oum,Kahn:2021ttr,Elor:2021swj,Arguelles:2022fqq,Alvey:2022pad,Su:2022wpj,PandaX:2023tfq,Su:2023zgr,Liang:2024xcx,Dutta:2024kuj,Bhattiprolu:2024dmh,Sun:2025gyj,Yang:2025ejt,Balan:2025uke,Gong:2025ves,Cheek:2025nul,Li:2025zwg,Ge:2025itf,Bernreuther:2025xqk,Hu:2025dsv,Cox:2025toz,Wang:2026you,Gong:2026dte}, such as semiconductor~\cite{SuperCDMS:2018mne, 2022PhRvD.106f2004A, CDEX:2023vvc, SENSEI:2023zdf, DAMIC-M:2025luv} and superfluid helium experiments~\cite{Schutz:2016tid,vonKrosigk:2022vnf,Hirschel:2023sbx,SPICE:2023tru,QUEST-DMC:2023nug}. However, to date, no unambiguous DM signal has been identified~\cite{XENON:2023cxc, PandaX:2024qfu, LZ:2024zvo,XENON:2025vwd}.

A recent high recoil energy analysis by the LUX–ZEPLIN (LZ) experiment has reportedly identified an event over the expected background in the region $E_{\rm nr} = 248 \pm 23$ keV~\cite{Akerib2026}. A localized feature at such high recoil energy is not characteristic of the monotonically falling spectrum expected from conventional elastic WIMPs--nucleus scattering. However, the inelastic DM scattering can instead select a specific recoil energy region through its threshold kinematics.

We consider an inelastic DM sector containing a ground state  $\chi$ and an excited state $\chi^\ast$, with masses $m_\chi$ and $m_{\chi^\ast}$, respectively~\cite{Tucker-Smith:2001myb,Tucker-Smith:2004mxa,Chang:2008gd,Cui:2009xq,Graham:2010ca,Barello:2014uda,Blennow:2015hzp,PandaX-II:2017zex,Baryakhtar:2020rwy,He:2020wjs,Choi:2020ysq,Bell:2021xff,Song:2021yar,Li:2022acp,XENON:2022avm,Eby:2023wem,Garcia:2024uwf}. The mass splitting is defined as $\delta \equiv m_{\chi^\ast}-m_\chi > 0$ and is assumed to be much smaller than the DM mass, $\delta \ll m_{\chi}$. The two possible scattering processes, $\chi+N\rightarrow\chi^\ast+N$ and $\chi^\ast+N\rightarrow\chi+N$, are endothermic and exothermic, respectively. The former consumes part of the incoming DM kinetic energy, whereas the latter converts the mass-splitting energy into additional kinetic energy of the final-state particles. In particular, a positive mass splitting raises the minimum incoming speed and confines the signal to a finite recoil energy interval~\cite{Bramante:2016rdh,PICO:2023uff,An:2025bby,Alloni:2026xdf}, making endothermic scattering particularly well suited to the LZ high recoil energy feature.

\begin{figure*}
    \centering
    \includegraphics[width=0.8\linewidth]{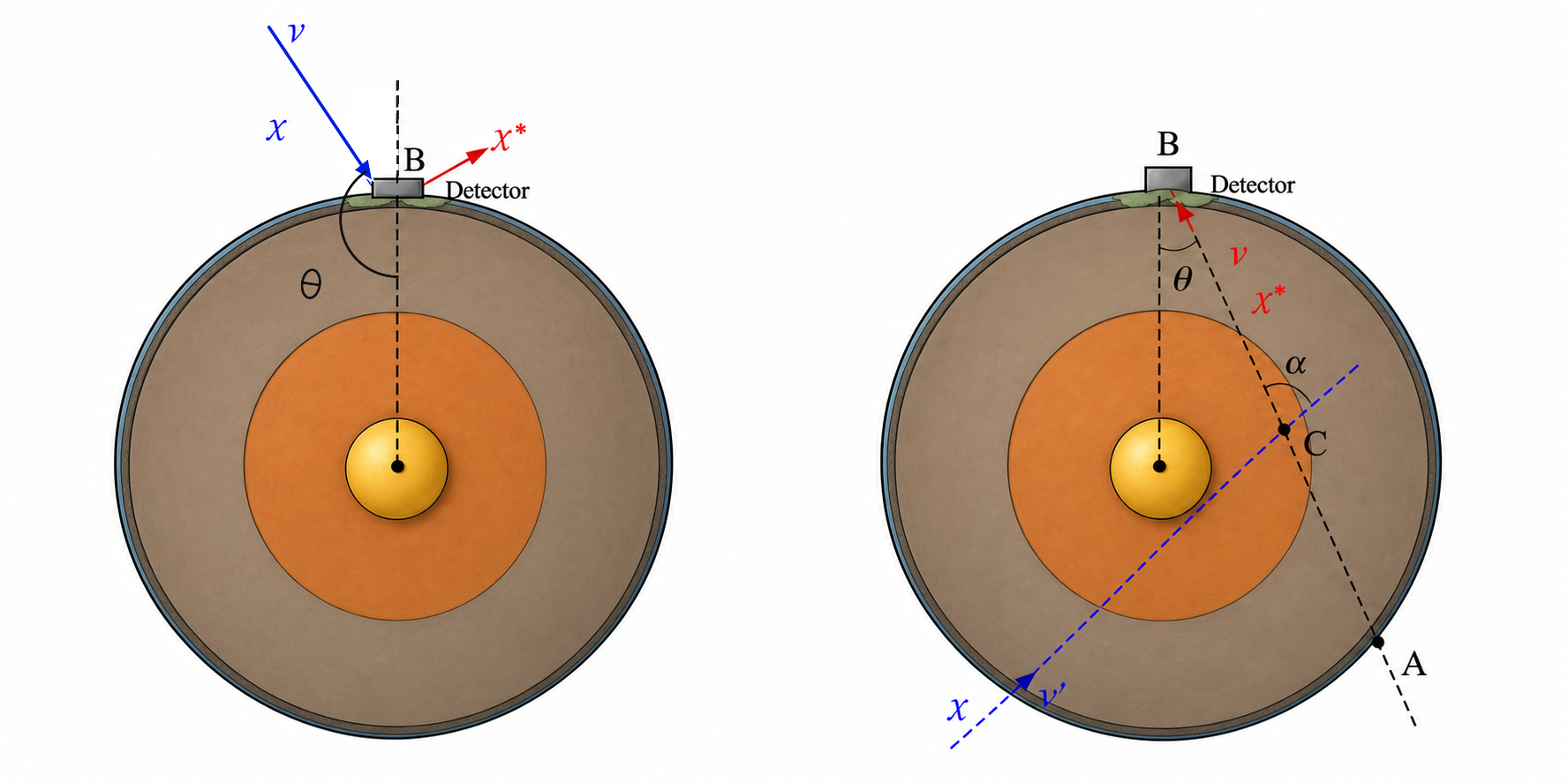}
    \caption{Schematic illustration of the two scattering processes. Left: a halo DM particle in the ground state $\chi$ reaches the detector directly and undergoes endothermic upscattering, $\chi + N\to\chi^\ast + N$. Right: an Earth-crossing particle $\chi$ first upscatters at $C$ inside the Earth and the resulting excited state $\chi^\ast$ propagates to the detector at $B$. The angles $\theta$ and $\alpha$ specify the incident trajectory and scattering geometry, respectively. The schematic was created with the assistance of OpenAI.}
    \label{fig:earth_scattering}
\end{figure*}

In this work, we assume that the ground state $\chi$ constitutes the dominant component of the DM halo in the Milky Way.  Therefore, two propagation contributions must be considered, as illustrated in Fig.~\ref{fig:earth_scattering}. For wind-facing directions, the terrestrial propagation distance is negligible compared with the DM mean free path, and ground-state halo DM can scatter directly from a target nucleus in the detector through the endothermic process. For Earth-crossing trajectories, propagation through the Earth may modify the incident flux. Within the single-scattering approximation, the flux arriving at the detector contains both an unscattered ground-state component and an excited-state component produced by one terrestrial upscattering. The surviving ground-state particles undergo the same endothermic process in the detector, whereas the Earth-produced excited states scatter exothermically through $\chi^\ast+N\to\chi+N$. In this work, we will calculate and compare the high recoil energy contributions from these two components, and perform a statistical analysis to determine the best-fit point in the $(m_\chi,\delta)$ plane at fixed $\bar{\sigma}_n$. Finally, we discuss the prospects for probing the best-fit point in other experiments, such as the planned upgrade of the Cryogenic Rare Event Search with
Superconducting Thermometers (CRESST) experiment~\cite{Angloher:2025fzw}.

\prlsection{Framework and Rate}{.} In this work, we assume that the inelastic DM interacts with Standard Model particles through a vector mediator $V_\mu$. The relevant interaction Lagrangian can be written as $\mathcal{L}_{\mathrm{int}} = g_\chi V_\mu\,\bar{\chi}^{\ast}\gamma^\mu\chi + g_q V_\mu \sum_q \bar q \gamma^\mu q + {\mathrm{H.c.}}$, where $g_\chi$ and $g_q$ denote the couplings of the vector mediator to the inelastic DM and to Standard Model quarks, respectively. For the endothermic process in the detector, the differential rate can be written as
\begin{equation}
    \frac{\mathrm{d} R}{\mathrm{~d} E_{\mathrm{nr}}}= N_T\frac{\rho_\chi}{ m_{\chi}} \int_{v>v_{\min }} v f(\mathbf{v}) \frac{\mathrm{d} \sigma}{\mathrm{~d} E_{\mathrm{nr}}} \mathrm{~d}^{3} \mathbf{v},
    \label{eq:rates}
\end{equation}
where $\rho_{\chi} = 0.3\; \mathrm{GeV/cm^3} $ is the local DM density, and $N_T$ is the number of target nuclei per unit detector mass. The differential cross section of DM-nucleus scattering can be given by 
\begin{equation}
    \frac{\mathrm{d} \sigma}{ \mathrm{d} E_{\rm nr}} = \frac{m_N \bar{\sigma}_{\mathrm{ n}}}{2 \mu_{\chi n}^2 v^2} [f_p Z +f_n(A-Z)]^2|F_{\rm DM}(E_{\rm nr})|^2 | F_N(E_{\rm nr})|^2,
\end{equation}
where $m_N$ and $\mu_{\chi n} = m_{\chi} m_n/(m_{\chi} +m_n)$ are the nucleus mass and the DM-nucleon reduced mass, respectively. The parameters $A$ and $Z$ are the mass and charge number, and $f_{p,n}$ are the dimensionless DM couplings to the proton and neutron, respectively. In this work, we assume $f_p =f_n=1$ for spin-independent DM-nucleus interaction. And the spin-independent DM-nucleon scattering cross section is defined as $\bar{\sigma}_n \equiv \mu_{\chi n}^2 \overline{|\mathcal{M}_n(q=0)|^2}/(16 \pi m_\chi^2 m_n^2)$ with the DM form factor $|F_{\mathrm{DM}}|^2 \equiv \overline{|\mathcal{M}_n(q)|^2}/\overline{|\mathcal{M}_n(q=0)|^2}$, where $\overline{|\mathcal{M}_n(q)|^2}$ is the matrix element for scattering on a free nucleon. In this work, we adopt $|F_{\mathrm{DM}}|^2 =1$ for a heavy mediator. The function $|F_{\mathrm{N}}(E_{\mathrm{nr}})|^2$ is the nuclear form factor, which can be described by the Helm form factor for the heavy nucleus. The minimum velocity $v_{\mathrm{min}}$ is required to produce a nuclear recoil energy $E_{\mathrm{nr}}$ for the mass splitting $\delta$,
\begin{equation}
    v_{\min }=\left|\frac{m_{N} E_{\mathrm{nr}}}{\mu_{\chi N}} \pm \delta\right| \frac{1}{\sqrt{2 E_{\mathrm{nr}} m_{N}}},
\label{eq:vmin}
\end{equation}
where $\mu_{\chi N}$ is the DM-nucelus reduced mass. The sign $+$ and $-$ denote the endothermic and exothermic processes, respectively.  

The differential rate formula, Eq.~\ref{eq:rates}, is valid for both the wind-facing and Earth-crossing directions. The difference lies in the DM velocity distribution, $f(\mathbf{v})$. For the wind-facing direction, the incoming DM flux originates directly from the Galactic halo, and its velocity distribution is described by the Standard Halo Model, $f_{\mathrm{SHM}}(\mathbf{v})$, assuming a Maxwell--Boltzmann distribution with halo velocity dispersion $\sigma_v = 156\;\mathrm{km/s}$, local escape speed $v_{\mathrm{esc}} = 544\;\mathrm{km/s}$, and Earth speed $v_e = 220\;\mathrm{km/s}$~\cite{Kavanagh:2016pyr}.

\begin{figure*}
    \centering
    \includegraphics[width=0.45\linewidth]{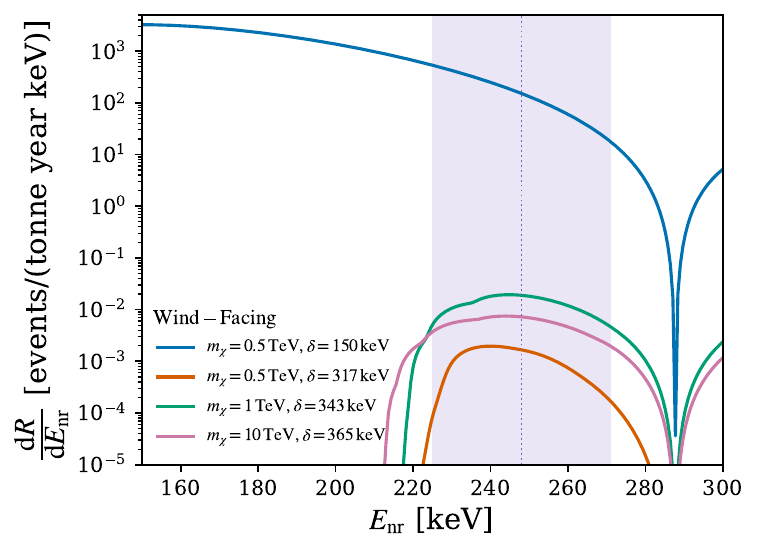}
    \includegraphics[width=0.45\linewidth]{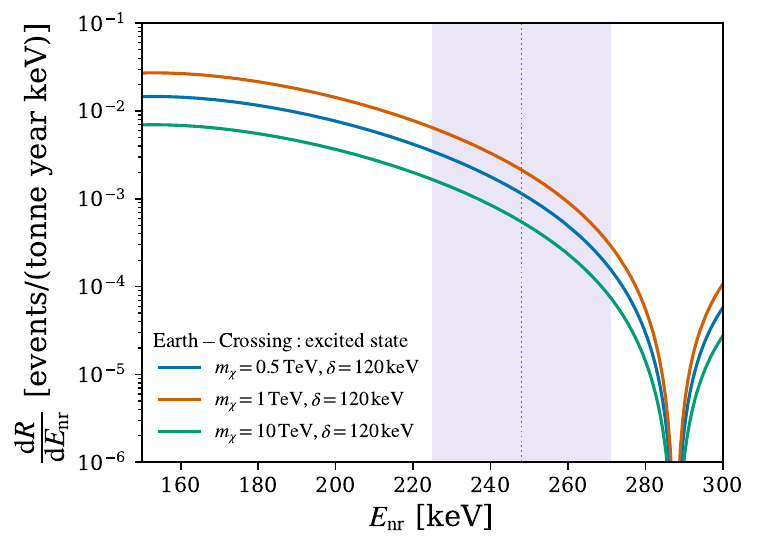}
    \caption{Differential nuclear recoil spectra for DM particles arriving from the ground--state in wind-facing (left) and excited state in Earth-crossing (right) directions, for the different benchmark DM masses and mass splittings. The shaded region denotes the LZ high recoil energy interval, $E_{\mathrm{nr}}\in[225,271]~\mathrm{keV}$, while the vertical dotted line marks its central value, $E_{\mathrm{nr}}=248~\mathrm{keV}$. Here the DM-nucleon scattering cross section is taken as $\bar{\sigma}_{n} = 10^{-38} \;\mathrm{cm}^2$.}
    \label{fig:rate}
\end{figure*}

For Earth-crossing directions, the standard DM flux is modified by DM--nucleus scattering as the DM particles propagate through the Earth~\cite{Kavanagh:2016pyr,Emken:2021vmf}. The velocity distribution of the excited state $\chi^\ast$ arriving at the detector can be written as~\cite{Emken:2021vmf}
\begin{equation}
\begin{aligned}
f^\ast(v)={}& \sum_{i,\pm} \int_0^1 \mathrm{d}\cos\theta \int_0^{2\pi}\mathrm{d}\phi \int_{-1}^{1}\mathrm{d}\cos\theta^\prime \int_{0}^{2 \pi} \mathrm{d} \phi^\prime\\
&\times \frac{\sigma_i\bar n_i d_{{\rm eff},i}(\cos\theta)}{2\pi}
\left| \frac{d\kappa_{\pm,i}^{-1}(v',\alpha)}{dv'} \right|^{-1}_{v'=\kappa_{\pm,i}(v,\alpha)} \\
&\times \frac{v'^3}{v}\, f_{\mathrm{SHM}}(v',\cos\theta', \phi^\prime)\, P_{\pm,i}(\cos\alpha).
\end{aligned}
\label{eq:excited_distribution}
\end{equation}
Here, $i$ labels the nuclear species inside the Earth, $\sigma_i$ is the total cross section for DM scattering on nucleus $i$. The $(v,\theta, \phi)$ and $(v^\prime, \theta^{\prime}, \phi^\prime)$ are the DM velocity after and before the scattering, and the scattering angle satisfies $\cos \alpha = \sin \theta \sin\theta^{\prime} \cos (\phi-\phi^{\prime}) + \cos \theta \cos \theta^{\prime}$, as shown in Fig.~\ref{fig:earth_scattering}. 

Moreover, the function $\kappa(v, \alpha)$ and its inverse $\kappa^{-1}(v^\prime, \alpha)$ describe the relation between $v$,$v^\prime$ and $\alpha$, i.e.,
\begin{equation}
    \begin{aligned}
    \kappa_{ \pm}(v, \alpha)&=v \frac{\cos \alpha \mp \sqrt{\frac{m_{N}^{2}}{m_{\chi}^{2}}-\sin ^{2} \alpha+\frac{2 \delta m_{N}\left(m_{N}-m_{\chi}\right)}{m_{\chi}^{3} v^{2}}}}{1-m_{N} / m_{\chi}}, \\
        \kappa_{ \pm}^{-1}\left(v^{\prime}, \alpha\right)&=v^{\prime} \frac{\cos \alpha \pm \sqrt{\frac{m_{N}^{2}}{m_{\chi}^{2}}-\sin ^{2} \alpha-\frac{2 \delta m_{N}\left(m_{N}+m_{\chi}\right)}{m_{\chi}^{3} v^{\prime 2}}}}{1+m_{N} / m_{\chi}},
    \end{aligned}
\end{equation}
where the sign $\pm$ denotes the different physical solutions, $\kappa_{\pm}>0$ and $\kappa^{-1}_{\pm}>0$. It is worth noting that only $\kappa_{+}$ solution is physical for $m_{\chi} <m_{N}$. The scattering probability with angle $\alpha$ is given by 
\begin{equation}
    P_{\pm}(\cos \alpha) = \left.\frac{1}{2} \frac{\dfrac{\mathrm{d} v}{\mathrm{d} \cos \alpha}\cos \alpha+v}{\sqrt{\frac{\mu_{\chi N}^2}{m_{\chi}^2}v^{\prime 2}-2 \frac{\mu_{\chi N}}{m_{\chi}^2}\delta}}\right|_{v= \kappa_{\pm}^{-1}(v^{\prime},\alpha)}.
\end{equation}

The effective Earth-crossing distance $d_{\mathrm{eff},i}$ is related to the number density $n_i(r)$ of nucleus $i$, which is defined by
\begin{equation}
    d_{\mathrm{eff},i}(\cos \theta) \equiv \int_{\mathrm{AB}} \mathrm{~d} l \frac{n_i(r)}{\bar{n}_i} \exp \left(-\frac{l}{v \tau}\right),
\end{equation}
where $\bar{n}_i \equiv \frac{1}{r_{\mathrm{E}}} \int_{0}^{r_{\mathrm{E}}}  n_i(r) \mathrm{~d} r$ is an averaged number density of nucleus $i$. The exponential weight factor $\exp \left(-\frac{l}{v \tau}\right)$ accounts for the decay of the excited state $\chi^*$ during its propagation to the detector, where $\tau$ is the mean lifetime of $\chi^{*}$. In this work, we assume that the lifetime of the excited state $\chi^{*}$ is sufficiently long that spontaneous de-excitation inside the Earth can be neglected, and we adopt $\tau=10^5~\mathrm{s}$ as a representative benchmark. Eq.~\ref{eq:excited_distribution} describes only the excited-state component produced by terrestrial upscattering. The surviving ground-state component is $f_{\chi}^{\mathrm{surv}}(\mathbf{v}) =P_{\mathrm{surv}}(\mathbf{v})f_{\mathrm{SHM}}(\mathbf{v})$, where $P_{\mathrm{surv}}(\mathbf{v}) =\exp[- \sum_i\int_{\mathrm{AB}}n_i(r)\sigma_i(v)\,\mathrm{d}l]$ is the survival probability.

In Fig~\ref{fig:rate}, we show the differential event rates induced by DM from the wind-facing (left panel) and Earth-crossing (right panel) directions. According to Eq.~\ref{eq:vmin}, the minimum and maximum nuclear recoil energy $E_{\mathrm{nr}}^{\pm} = \frac{\mu_{\chi N}^{2}}{2m_N}\left[v_{\max}\pm\sqrt{v_{\max}^{2}-\frac{2\delta}{\mu_{\chi N}}}\right]^2$ subject to $\delta\leq\mu_{\chi N}v_{\max}^{2}/2$, where $v_{\max}$ is the maximum DM speed. The blue and orange lines in the left panel demonstrate that a smaller mass splitting broadens the kinematically accessible recoil energy, while its upper endpoint increases with the DM mass over the range considered. Explaining the LZ spectrum in $E_{\mathrm{nr}} \in[225,271]$ keV requires a DM mass larger than approximately 500~GeV and a corresponding mass splitting of $\mathcal{O}(300)~\mathrm{keV}$. The Earth-produced excited-state component cannot account for this
feature. Because the nuclei abundant in the Earth are lighter than xenon, terrestrial upscattering is subject to the more restrictive condition $\delta\leq\mu_{\chi A}v_{\max}^{2}/2$, where $A$ denotes a terrestrial nuclear species. For a mass splitting of $\mathcal{O}(300)~\mathrm{keV}$, the endothermic transition $\chi+N_A\to\chi^\ast+N_A$ inside the Earth is kinematically forbidden or strongly suppressed. Consequently, the Earth-produced excited-state component required for subsequent exothermic scattering in the detector is negligible, as illustrated in the right panel of Fig.~\ref{fig:rate}. In this limit, however, the Earth becomes transparent to the ground-state component, such that $P_{\mathrm{surv}}\to1$. These surviving ground-state particles can still reach the detector and undergo direct endothermic scattering on xenon. This phenomenon will eliminate the daily modulation from the wind-facing direction.

\prlsection{Results}{.} To better characterize the overall behavior of the nuclear recoil energy spectrum, we include both the low and high energy regions in our statistical analysis. In this work, we perform a two-bin analysis covering the intervals $E_{\mathrm{nr}}\in[160,200]~\mathrm{keV}$ and $E_{\mathrm{nr}}\in[225,271]~\mathrm{keV}$. The corresponding observed event counts are taken to be $n_1=0$ and $n_2=1$, respectively.  The predicted number of signal events in the $i$-th energy bin is
\begin{equation}
    \mathcal{N}_i = \mathcal{E} \int_{E_{\mathrm{nr},i}^{\min}}^{E_{\mathrm{nr},i}^{\max}} \mathrm{d} E_{\mathrm{nr}} \frac{\mathrm{d} R}{\mathrm{d} E_{\mathrm{nr}}} \epsilon(E_{\mathrm{nr}}),
\end{equation}
where $\mathcal{E}=2.84~\mathrm{ton\cdot yr}$ is the effective exposure and $\epsilon$ is the detection efficiency with nuclear recoil energy $E_{\mathrm{nr}}$ of the LZ experiment. Treating the two bins as statistically independent Poisson counting experiments and neglecting background and its uncertainties, the likelihood is given by 
\begin{equation}
\mathcal{L}(m_\chi,\delta,\bar{\sigma}_n)=\prod_{i=1}^{2}\frac{\mathcal{N}_i^{n_i}e^{-\mathcal{N}_i}}{n_i!}.
\end{equation}
The associated likelihood-ratio statistic can be expressed as $\chi^2\equiv-2\ln\mathcal{L}$, and $\Delta\chi^2\equiv\chi^2-\chi^2_{\min}$, where $\chi^2_{\min}$ is the minimum value over the parameter space. The parameter $\Delta\chi^2$ measures the deviation from the best-fit point and is used to determine the corresponding confidence regions. 

\begin{figure}
    \centering
    \includegraphics[width=1\linewidth]{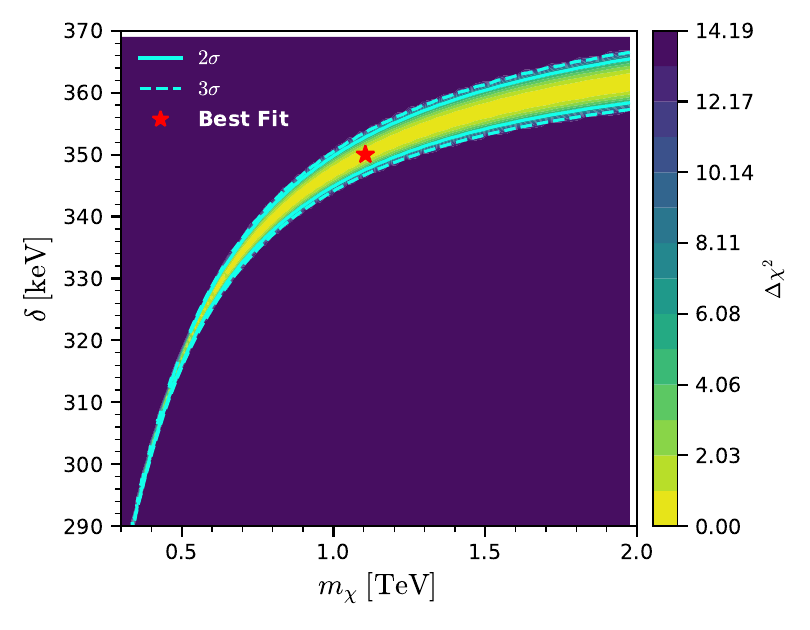}
    \caption{Distribution of $\Delta\chi^2$ in the $(m_\chi,\delta)$ plane for the fixed cross section $\bar{\sigma}_n=10^{-37}~\mathrm{cm}^2$. The red star marks the best-fit point, $(m_\chi,\delta)\simeq(1.105~\mathrm{TeV},350~\mathrm{keV})$, while the light-blue solid and dashed contours indicate the $2\sigma$ and $3\sigma$ likelihood contours, respectively.}
    \label{fig:bestfit}
\end{figure}

In this work, we employ the \texttt{pyhf} package~\cite{pyhf,pyhf_joss} to scan the parameter region $300~\mathrm{GeV}\leq m_\chi\leq2~\mathrm{TeV}$ and $290~\mathrm{keV}\leq\delta\leq370~\mathrm{keV}$ for a fixed cross section $\bar{\sigma_n}=10^{-37}~\mathrm{cm}^2$. The best-fit point and the corresponding $2\sigma$ and $3\sigma$ likelihood contours are determined from the resulting $\Delta\chi^2$ distribution. As shown in Fig.~\ref{fig:bestfit}, the best-fit point is located at $m_\chi\simeq1.105~\mathrm{TeV}$ and $\delta\simeq350~\mathrm{keV}$. The boundaries of the $2\sigma$ and $3\sigma$ regions are indicated by the light-blue solid and dashed contours, respectively. The precise location of the best-fit point may depend on the number and placement of the energy bins. A finer binning retains more information about the spectral shape and generally leads to a more stable determination of the preferred parameters. The assumed fixed cross section can also shift the best-fit point. Therefore, we repeat the scan over the same $(m_\chi,\delta)$ parameter space for $\bar{\sigma}_n=10^{-38}~\mathrm{cm}^2$ and $10^{-36}~\mathrm{cm}^2$. The corresponding best-fit points are $(m_\chi,\delta)\simeq(0.687~\mathrm{TeV},331~\mathrm{keV})$ and $(1.942~\mathrm{TeV},364~\mathrm{keV})$, respectively. For $\bar{\sigma}_n=10^{-37}~\mathrm{cm^2}$, in addition to the nominal best-fit point $(m_\chi,\delta)=(1.105~\mathrm{TeV},350~\mathrm{keV})$, three other points, $(0.994~\mathrm{TeV},347~\mathrm{keV})$, $(0.576~\mathrm{TeV},325~\mathrm{keV})$, and $(1.778~\mathrm{TeV},360~\mathrm{keV})$, satisfy $\Delta\chi^2<10^{-4}$. These points predict event numbers close to $(\mathcal{N}_{\mathrm{1}},\mathcal{N}_{\mathrm{2}})=(0,1)$ and therefore fit the two-bin data nearly equally well. In the following discussion, we therefore adopt the best-fit point obtained for $\bar{\sigma}_n=10^{-37}~\mathrm{cm}^2$ as our reference benchmark.

\begin{figure}
    \centering
    \includegraphics[width=1\linewidth]{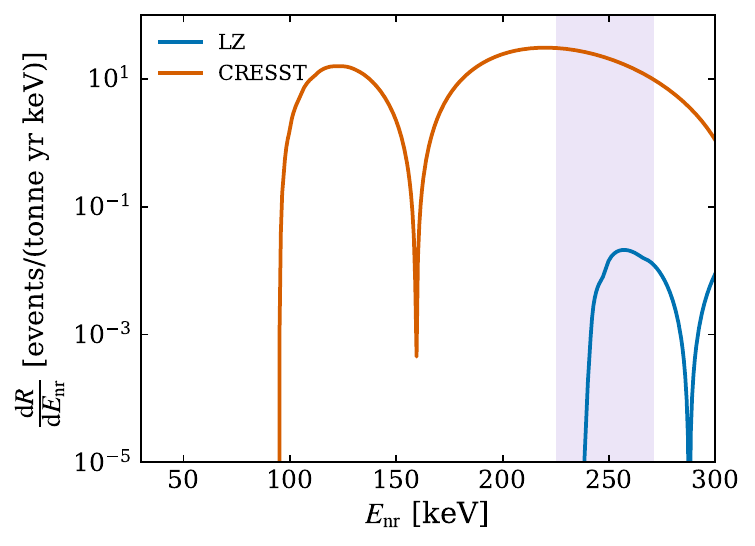}
    \caption{Differential event rates in LZ (blue) and CRESST (orange) for the best-fit inelastic DM parameters, $(m_\chi,\delta)=(1.105~\mathrm{TeV},350~\mathrm{keV})$, with $\bar{\sigma}_n=10^{-37}~\mathrm{cm}^2$.}
    \label{fig:LZ_CRESST}
\end{figure}
For the large mass splitting associated with the best-fit point, endothermic upscattering is kinematically inaccessible or strongly suppressed in many DM and neutrino experiments employing target nuclei lighter than xenon. In contrast, the tungsten nuclei in the \(\mathrm{CaWO}_4\) targets of CRESST are heavier than xenon and allow the same transition to occur at lower nuclear recoil energies. In Fig.~\ref{fig:LZ_CRESST}, we compare the differential event rates expected in LZ and CRESST at the best-fit point, retaining the contribution of tungsten and assuming unit detection efficiency for CRESST. Although the CRESST-II and CRESST-III searches have primarily focused on low recoil energies, the CRESST-II data were displayed up to approximately \(120~\mathrm{keV}\)~\cite{CRESST:2015txj}, demonstrating that energy depositions can be reconstructed in this range. Moreover, the schematic light-yield distribution presented in the official CRESST detector description shows the tungsten-recoil band extending to $150~\mathrm{keV}$~\cite{CRESSTDetectorWebsite}. We therefore adopt $E_{\mathrm{nr}}^{\max}=150~\mathrm{keV}$ as an illustrative upper limit for a prospective high-energy CRESST analysis. 

At the best-fit point and for $\bar{\sigma}_n=10^{-37}~\mathrm{cm^2}$, the integrated recoil rate over $[95,150]~\mathrm{keV}$ is $R_{95-150}\simeq
0.515~\mathrm{events\cdot 
kg^{-1} \cdot yr^{-1}}$. The corresponding expected event is
\begin{equation}
    \mathcal{N}_{95-150}=0.515\left(\frac{\bar{\sigma}_n}{10^{-37}~\mathrm{cm^2}}\right)\left(\frac{\mathcal E}{\mathrm{kg\cdot yr}}\right)\bar{\epsilon},
\end{equation}
where $\bar{\epsilon}$ is the rate-weighted detection efficiency.  For reference, in a background-free Poisson counting experiment with zero observed events, the 90\% C.L. upper limit corresponds to 2.3 signal events.  Assuming unit efficiency, obtaining $\mathcal{N}_{95-150}=2.3$ would require an exposure of approximately $4.47~\mathrm{kg\cdot yr}$, or equivalently $1.63~\mathrm{tonne\cdot day}$.
The planned CRESST upgrade comprises approximately 100 detectors, with a representative configuration of 70 $24~\mathrm{g}$ and 26 $2~\mathrm{g}$ $\mathrm{CaWO}_4$ modules, corresponding to a total target mass of approximately $1.73~\mathrm{kg}$~\cite{Angloher:2025fzw}. The projected exposure is approximately $500~\mathrm{kg\cdot day}$ after one year and may reach $1.5~\mathrm{ton\cdot day}$ after three years, which is close to the $1.63~\mathrm{ton\cdot day}$ required in our benchmark scenario. The predicted signal could therefore be within reach of the upgraded CRESST experiment and may become observable with a modest extension of the data-taking period. A dedicated analysis of the detector response, signal efficiency, nuclear-recoil acceptance, and background level in the $95-150~\mathrm{keV}$ interval would further establish the sensitivity to this signal.

\prlsection{Conclusion}{.} In this work, we showed that endothermic inelastic dark matter can account for the high recoil energy feature reported by the LUX--ZEPLIN Collaboration. The parameter region capable of reproducing the spectrum is characterized by $m_\chi\gtrsim500~\mathrm{GeV}$ and $\delta=\mathcal{O}(300)~\mathrm{keV}$. For such a large splitting, upscattering on the nuclei abundant in the Earth is kinematically forbidden, suppressing the Earth-produced excited-state component. Ground-state particles can nevertheless traverse the Earth without scattering and subsequently undergo direct endothermic scattering on xenon. An illustrative two-bin likelihood analysis identifies $(m_\chi,\delta)\simeq(1.105~\mathrm{TeV},350~\mathrm{keV})$ as a representative benchmark for $\bar{\sigma}_n=10^{-37}~\mathrm{cm^2}$. Future LZ data will determine whether the reported event is a statistical fluctuation or a possible signature of inelastic dark matter, while the planned CRESST upgrade provides a complementary test of the preferred parameter region.

\section*{Acknowledgements}
This work was supported by the Alexander von Humboldt Foundation, by the National Natural Science Foundation of China (NNSFC) under grant No.12335005, and by the PI Research Fund (Grant No. 5101029470335) from Henan Normal University.

\bibliography{refs}

\end{document}